\documentclass[preprint,journal]{vgtc}            

\newcommand*{\anj}{\textcolor{black}}

\onlineid{1726}

\vgtccategory{Research}

\vgtcpapertype{Evaluation}

\title{Mind Drifts, Data Shifts: Utilizing Mind Wandering to Track the Evolution of User Experience with Data Visualizations }

\newcommand\sbullet[1][.5]{\mathbin{\vcenter{\hbox{\scalebox{#1}{$\bullet$}}}}}

\author{Anjana Arunkumar, Lace Padilla, and Chris Bryan}
\authorfooter{
$\sbullet$ Anjana Arunkumar and Chris Bryan are with Arizona State University. Email: aarunku5@asu.edu, cbryan16@asu.edu.\\
$\sbullet$ Lace Padilla is with Northeastern University, Email: l.padilla@northeastern.edu.
}

\vspace{-5em}
\abstract{%
User experience in data visualization is typically assessed through post-viewing self-reports, but these overlook the dynamic cognitive processes during interaction. This study explores the use of mind wandering\anj{-- a phenomenon where attention spontaneously shifts from a primary task to internal, task-related thoughts or unrelated distractions--} as a dynamic measure during visualization exploration. Participants reported mind wandering while viewing visualizations from a pre-labeled visualization database and then provided quantitative ratings of trust, engagement, and design quality, along with qualitative descriptions and short-term/long-term recall assessments. Results show that mind wandering negatively affects short-term visualization recall and various post-viewing measures, particularly for visualizations with little text annotation. Further, the type of mind wandering impacts engagement and emotional response. \anj{Mind wandering also functions as an intermediate process linking visualization design elements to  post-viewing measures, influencing how viewers engage with and interpret visual information over time.} Overall, this research underscores the importance of incorporating mind wandering as a dynamic measure in visualization design and evaluation, offering novel avenues for enhancing user engagement and comprehension.

}


\teaser{
  \centering
  \includegraphics[width=0.7\linewidth]{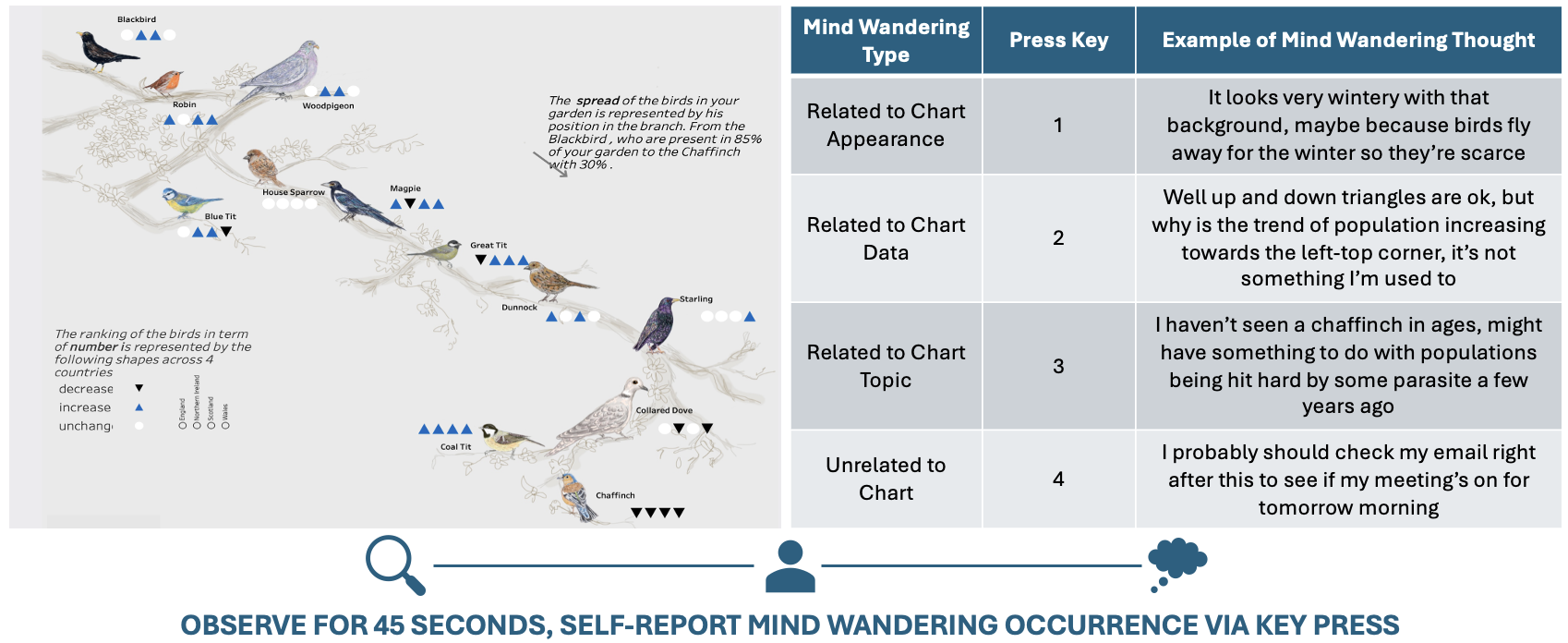}
  \caption{%
\anj{Types of mind wandering self-reported via key press by participants over a 45 second period, when viewing a data visualization. Mind wandering can be relevant in nature, by still focusing on a visualization (1) aesthetically, (2) cognitively, or (3) affectively. Alternately, it can be (4) completely unrelated to the visualization.} 
  }
  \label{fig:teaser}
}

\graphicspath{{figs/}{figures/}{pictures/}{images/}{./}} 

\usepackage{tabu}                      
\usepackage{booktabs}                  
\usepackage{lipsum}                    
\usepackage{mwe}                       
\usepackage{mathptmx}                  
\usepackage{soul}
\usepackage{graphicx}
\usepackage{float}
\usepackage{multirow}

\begin{document}

\maketitle

\section{Introduction}

Data visualization has become an integral part of modern information communication, serving as a potent tool for conveying complex data and insights to users~\cite{9222102}. Yet, despite its ubiquity, the understanding of user experience in data visualization remains multifaceted and challenging to grasp~\cite{bertini2020shouldn,lee2022affective}. Traditionally, user experience evaluation in this domain has relied heavily on post-viewing self-reports, capturing static impressions of trust~\cite{elhamdadi2023vistrust,pandey2023you}, emotion~\cite{10294209,lan2023affective}, aesthetics~\cite{he2022beauvis, bateman2010useful}, design-quality~\cite{xiong2021visual,arunkumar2021bayesian}, memorability~\cite{borkin2013makes,borkin2015beyond}, engagement~\cite{mahyar2015towards,boy2015storytelling,haroz2015isotype}, etc. However, such approaches often overlook the dynamic nature of cognitive processes that unfold during interaction with visualizations. 

To disentangle these confounding factors, we set out to answer a fundamental question: \textit{How can we effectively measure the occurrence and impact of fluctuations in user cognition during the observation of data visualizations, while ensuring a realistic user experience?} Prior work in visual cognition literature has extensively studied the impact of mind-wandering-- a phenomenon characterized by shifts in attention away from the primary task or stimuli towards internal thoughts or unrelated external stimuli-- on various cognitive processes~\cite{faber2020eye}. This includes investigations into its effects on perceptual sensitivity~\cite{zhang2021refixation,zhang2022perceptual}, memory encoding and retrieval~\cite{blonde2022wandering,soemer2020working,taatgen2021resource}, as well as visual task performance~\cite{maillet2020age,brosowsky2020mind}. Additionally, recent studies have elucidated the intricate interplay between mind-wandering and emotional states~\cite{pelagatti2020closer,banks2022individual}. Building upon this existing research, we \anj{posit that mind-wandering can be leveraged as a dynamic measure to track shifts in the nature and focus of a user's experience~\cite{walny2017active,hullman2011benefitting} during visualization viewing}, and explore how it may reflect in post-viewing emotional responses and engagement levels.  

In this work, we utilized a taxonomized dataset of 100 static, real-world visualizations~\cite{10294209}, encompassing a range of aesthetics, to conduct a controlled experiment to understand how mind-wandering influences users' perceptions and interactions with visual data. \anj{For 50 stimuli, participants first report at-a-glance comprehension to establish a baseline for understanding, followed by reports of different types of mind-wandering occurrences over a 45 second period using established methods~\cite{seli2013few,kane2021testing}, as shown in Fig.~\ref{fig:teaser}. In order to validate mind wandering as a dynamic measure, we examined its impact on post-viewing reports of overall cognitive and affective enagement, as well as constructs that reflect the implications of engagement patterns like trust, design quality, and emotional response, the verbalized recall of key details, and a visual recognition test one week later for retention assessment.}

\anj{At a high level, the study results indicate several interesting findings, such as that mind-wandering, regardless of its relevance to the chart topic, impaired short-term recall, and diminished viewer trust and design-quality ratings. The presence of certain visualization elements also correlated with increased instances of mind wandering, and mind-wandering could mediate the relationship between visualization encoding features and collected post-viewing measures.} 

\textbf{Contributions:} This work pioneers the investigation of in situ user experience during data visualization viewing, going beyond traditional post-viewing static assessments. To promote reproducibility, all study materials, visualization labeling and metadata, participant demographics, analysis, and results are \anj{publicly available at: \url{https://osf.io/h5awt/}}. Our experiment yields valuable insights into the dynamic nature of user experience, highlighting the impact of mind-wandering on viewer ratings, and identifying key design factors influencing viewer perceptions and behaviors in real-time. These findings emphasize the importance of considering in situ user experience in visualization design and evaluation, contributing to the advancement of dynamic evaluation methodologies in the field.
\section{Background}

\subsection{Mind Wandering and Visual Cognition}

Mind wandering is when attention drifts from the current task to internally generated thoughts, memories, or fantasies~\cite{smallwood2015mind,christoff2016mind}. It often occurs spontaneously and unconsciously~\cite{smeekens2016working,dorsch2015focused}. While it can sometimes be beneficial, leading to creativity or problem-solving, it can also impair tasks requiring sustained focus~\cite{zedelius2018unraveling}. Mind wandering isn't always a sign of disengagement; it serves various cognitive functions with both positive and negative outcomes. Relevant mind wandering aligns with current goals, like contemplating related projects during a meeting~\cite{smallwood2013not,vannucci2019thought}. In contrast, irrelevant mind wandering, such as daydreaming during a lecture, disrupts performance~\cite{blonde2022wandering,blonde2022bored}. 

Researchers have explored mind wandering in diverse settings, including reading tasks, visual tasks, driving, problem-solving, and computer-based activities~\cite{feng2013mind,yanko2014driving,krimsky2017influence}. To measure it, studies have employed various methods such as self-report questionnaires~\cite{seli2013few}, thought sampling~\cite{smith2018mind}, task performance analysis~\cite{thomson2014link}, and physiological measures like heart rate variability, pupil dilation, or brain activity (e.g., EEG)~\cite{pelagatti2020closer}. \anj{In this work, we utilize a key-press paradigm, drawing from literature where participants are asked to report if they are on-task vs. mind-wandering~\cite{thomson2014link} or on the nature of mind-wadnering (i.e., relevant vs. irrelevant)~\cite{seli2018role}.} 

The relationship between mind wandering and visual cognition is multifaceted. Mind wandering involves internally generated thoughts, often intertwined with visual imagery, which can significantly impact cognitive processes~\cite{shrimpton2017daydream}. For instance, individuals may mentally visualize scenes or images associated with their thoughts during daydreaming or memory recall. Moreover, external visual stimuli can trigger mind wandering by capturing attention, leading to a cascade of internally generated thoughts, whether relevant or unrelated to the task at hand~\cite{shrimpton2017daydream,schmitt2023mind}. Previous research has shown that mind wandering during visual processing can disrupt memory consolidation and decrease the accuracy of recall for visual stimuli~\cite{nicosia2022targeted,blonde2022wandering}.

Understanding the interplay between mind wandering and visual cognition is crucial for comprehensively understanding how viewers experience data visualization exploration. Hence, this work emphasizes the importance of considering cognitive dynamics in the design of data visualizations and to the enhancement of user experiences in data-driven domains. This holistic approach not only enriches our understanding of visualization cognition but also informs the development of more effective interventions aimed at optimizing cognitive processes during visual information processing. It's noteworthy to mention that despite the extensive study of mind wandering in various aspects of vision and visual stimuli, there remains a gap in research concerning its interplay with data visualizations.

\subsection{Perception and Memorability of Visualizations}
\anj{
Studies on the perception of visualizations investigate how they are perceptually and comprehensively processed~\cite{healey2012attention}. Factors such as the types of visual encodings used, the data dimensions being shown~\cite{munzner2014visualization,padilla2020powerful}, the data volume displayed~\cite{keim2013big}, the design stylings chosen~\cite{kim2016data}, the presence of narrative elements within the visualization~\cite{hullman2011visualization}, and the viewer's cognitive focus~\cite{healey2012attention} influence chart readability, as well as insights gleaned perceptually. Previous research has aimed to understand how these design choices impact memorability and comprehensibility across various types of visualizations~\cite{hullman2011benefitting,few2020chartjunk,tufte1985visual}. These aspects have been emphasized in the ``slow analytics'' movement, which emphasizes the importance of understanding and retaining analytical tasks rather than solely focusing on precision~\cite{bradley2021approaching,lupi2017data}, by embellishing visualizations to introduce ``visual difficulties" through the use of human-recognizable objects and color~\cite{borkin2013makes,borkin2015beyond}.
}

In this study, we explore the dynamic nature of visual cognition during visualization consumption, moving beyond mere perception. Specifically, we investigate how individuals dynamically process visualizations for interpretation and how this cognition process influences their ability to recall information in both the short and long term. By investigating the impact of mind wandering on the accuracy and focus of recall, we provide insights that can inform the design of visualizations that promote sustained attention.

\begin{figure*}[!th]
    \centering
    \includegraphics[width=0.7\textwidth]{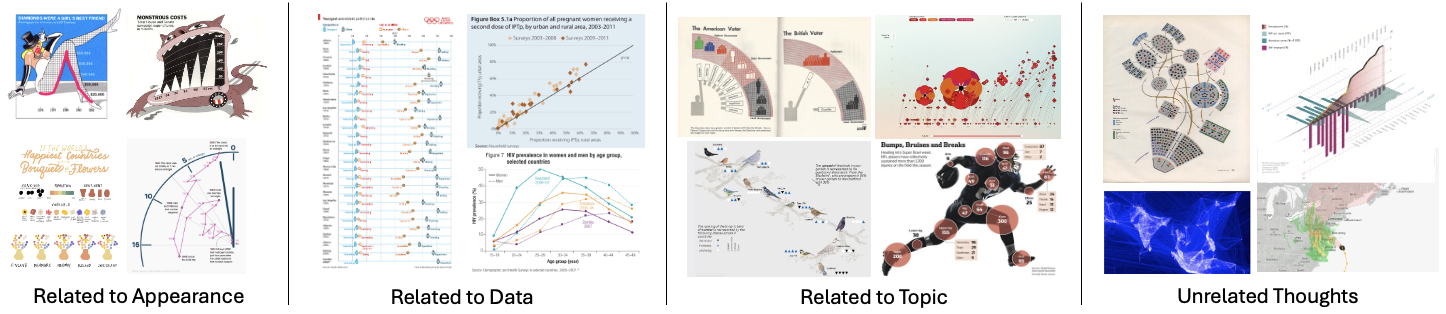}
    \caption{\anj{Examples of visualization stimuli~\cite{10294209} that trigger early and more frequent reports of different types of mind-wandering.}}
    \label{fig:stimuli}
    \vspace{-1em}
\end{figure*}

\subsection{Visualization Engagement}

\anj{
\textit{User engagement} has been defined in various ways by the visualization community. Application-based evaluation often centers on users' exploration efforts~\cite{haroz2015isotype,boy2015storytelling}, with some consideration of user's goals, aesthetic preferences, data familiarity, and other relevant display parameters~\cite{yi2008understanding}. Such factors have been found to influence users' overall exploration interest and the quality of insights extracted~\cite{mahyar2015towards,haroz2015isotype}. These are typically collected as post-viewing measures, once the user has been exposed to the visualization stimulus. Additional measures collected in situ viewing include the monitoring of mouse hover and clicks~\cite{boy2015storytelling,haroz2015isotype}; while these measures have been used to reflect on user attention, considerations of pauses in a user's cognitive process due to attentional lapses have not been examined. In this context, we propose that mind wandering provides a richer measure of engagement. Relevant mind wandering may indicate a shift from higher-level cognitive tasks, suggesting balanced engagement, while focused attention may indicate active involvement in one aspect. In order to better quantify the nature and level of engagement, we juxtapose mind-wandering metrics collected in our study against traditional post-viewing measures of engagement~\cite{bloom2020taxonomy}. For cognitive engagement, Mahyar et al.\cite{mahyar2015towards} synthesize a five-level taxonomy (Expose-Involve-Analyze-Synthesize-Decide) which suggests that engagement increases as users perform higher-level cognitive tasks. Lee-Robbins et al. \cite{lee2022affective} complement this with affective engagement objectives (Observe-Position-Strengthen-Connect-Behave) aiming to influence audience opinions, attitudes, or values.
}

\anj{However, it is essential to recognize that mind wandering is not considered as a definitive, all-encompassing measure of engagement. Rather, it is held as an indicator that could potentially complement other metrics like user goals, familiarity with data, and interaction styles, which have been identified as significantly reflecting the nuances of engagement in prior work. By incorporating these perspectives, we can use mind wandering as a dynamic measure to develop a more nuanced understanding of post-viewing engagement self-reports by users.}

\subsection{Dynamic Measures for Visualization User Experience}

Dynamic measurement of user experience in data visualization involves capturing real-time cognitive and behavioral interactions to understand how individuals perceive, interpret, and engage with visual information, enhancing our understanding of user engagement and decision-making processes~\cite{willigen2019measuring,alexander2021improving}. For instance, eye tracking offers insights into how people focus on data visualizations, serving as a proxy for visual attention~\cite{matzen2017patterns}. This data can be used to construct visual saliency models, predicting which chart areas attract attention~\cite{5963660}. While eye tracking focuses on perception, mind wandering reveals task-based cognition, offering insights into how individuals mentally engage with and interpret visual information beyond perception.

Think-aloud protocols, where viewers narrate their reasoning in real-time while viewing stimuli, are another form of dynamic cognitive interview\cite{reinhart2022think}. This method helps researchers understand the underlying reasoning behind participants' actions, uncovering patterns, preferences, and areas of confusion, along with spontaneous reactions and emotional responses, revealing participants' subjective experiences and perceptions of the data visualization\cite{fan2023understanding,blascheck2016triangulating}. In contrast, while mind wandering may provide less information-rich data during sustained attention tasks, it offers a more naturalistic representation of cognitive processes compared to think-aloud protocols\cite{jordano2018often}. When individuals continuously verbalize their thoughts, they may become more self-conscious and deliberate\cite{reinhart2022think}, potentially altering their cognitive processes and decision-making strategies.

To the best of our knowledge, this work represents the first attempt to reconcile post-viewing reports, such as overall trust, engagement, design quality, etc., with a dynamic mind wandering measure in the context of visualizations. While previous research has often linked dynamic measures to task performance, judgment, or decision quality, our approach offers a more comprehensive understanding of user experience by integrating subjective post-viewing assessments with real-time cognitive processes. 

\begin{figure*}[!ht]
  \centering
  \includegraphics[width=0.75\textwidth]{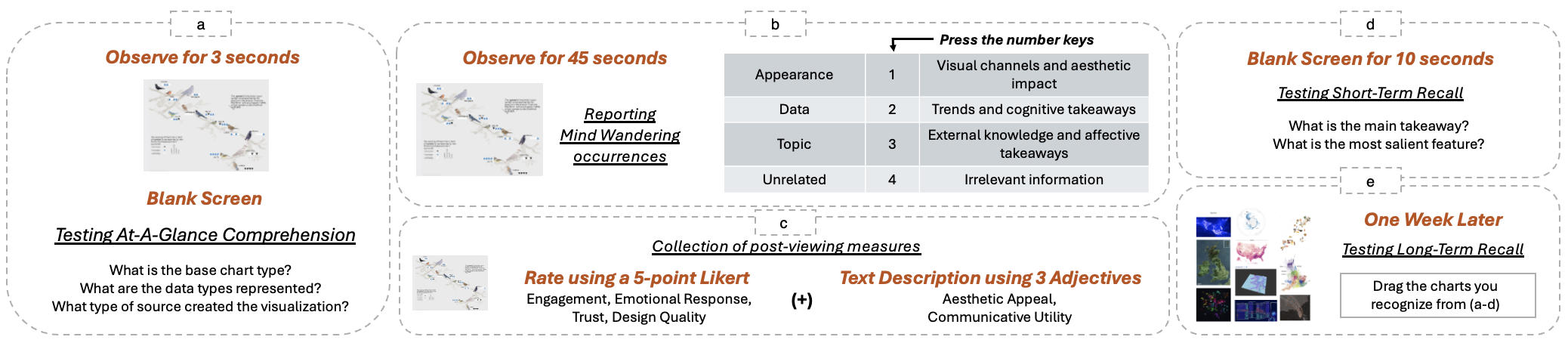}
    \caption{Study procedure: (a) Phase 1: 50 trials, tests at-a-glance comprehension. (b--d) Phase 2: 50 trials, mind-wandering self-reports~\cite{d2016attending}, collect post-viewing measures, short-term recall. (e) Phase 3: 1 trial, long-term recall tested with a visual recognition task.}
    \label{fig:survey}
\end{figure*}

\begin{table*}[!htb]
\centering
\caption{\anj{Qualitatively coding adjective valence and strength based on semantic similarity to existing adjective corpii.}}
\resizebox{0.75\textwidth}{!}{%
\begin{tabular}{@{}llllll@{}}
\toprule
\multicolumn{1}{c}{\multirow{2}{*}{\textbf{Strength}}} & \multicolumn{1}{c}{\multirow{2}{*}{\textbf{Adverbial Modifier}}} & \multicolumn{2}{c}{\textbf{Aesthetic Appeal}} & \multicolumn{2}{c}{\textbf{Communicative Utility}} \\ \cmidrule(l){3-6} 
\multicolumn{1}{c}{} & \multicolumn{1}{c}{} & \multicolumn{1}{c}{\textbf{Positive Valence}} & \multicolumn{1}{c}{\textbf{Negative Valence}} & \multicolumn{1}{c}{\textbf{Positive Valence}} & \multicolumn{1}{c}{\textbf{Negative Valence}} \\ 
\midrule
1 : Very Low Intensity & extreme/absolute & pleasant, simple & plain, basic & clear, simple  & confusing, unclear \\
2 : Low Intensity & very/rather & clean, attractive & dull, ordinary & understandable, informative & misleading, vague \\
3 : Lower Medium Intensity & good/decent & elegant, charming & mediocre, uninspiring & effective, useful &  ambiguous, ineffective\\
4 : Upper Medium Intensity & good/decent & striking, captivating & cluttered, chaotic & insightful, persuasive & incoherent, unclear \\
5 : High Intensity & very/rather & stunning, gorgeous & disorganized, unappealing & impactful, compelling & deceptive, misinterpreted \\
6 : Very High Intensity & extreme/absolute & breath-taking, exquisite & repulsive, hideous & transformative, enlightening & irrelevant, misguided \\ \bottomrule
\end{tabular}%
}
    \vspace{-1em}
\label{tab:2}
\end{table*}

\section{Study}

\subsection{Research Questions}

We explored three primary research questions, outlined below. Our analysis, detailed in Section~\ref{sec:5}, is structured around these questions and their corresponding hypotheses. Due to the exploratory nature of this study, we did not make specific predictions regarding the strength or direction of the hypothesized effects.

\noindent\textbf{RQ1: }Does mind wandering during visualization exploration influence viewer experience? \anj{Our initial hypothesis was that mind wandering would influence both cognitive and affective visualization engagement.} \textbf{RQ2: }How do visualization design elements contribute to the frequency and temporal distribution of mind wandering during visualization exploration? \anj{We hypothesized that different visualization elements would have varying effects on the occurrence of mind wandering. We further hypothesized that elements linked with barriers to visualization comprehension, such as visual density, data volume, and encoding dimensionality would lead to increased mind wandering.} 
\\\textbf{RQ3: } Does mind wandering serve as a mediator between visualization encoding features and user-reported post-viewing measures? \anj{Engagement has been previously studied as a moderator in the influence of visualizations on post-viewing task measures~\cite{sanchez2023effect}. We propose a sequential process where visualizations initially affect mind wandering, which in turn influences engagement levels, thereby affecting variations in reported post-viewing measures. Hence we hypothesize that that both the frequency and temporal distribution of mind wandering act as mediators linking design elements to the outcome measures.}

\subsection{Stimuli Overview}
\label{sec:3}

Communicative visualizations~\cite{card1999using} constitute the primary exposure individuals have to visualizations~\cite{9222102}. Beyond the visualization community, infographics are frequently employed to enhance audience engagement~\cite{houts2006role} and comprehension of information~\cite{dick2014interactive}, particularly for individuals with low-to-medium graph literacy~\cite{burns2021designing}. We gathered a dataset of 100 visualizations curated by Arunkumar et al.~\cite{10294209} (see a subset in Fig.~\ref{fig:stimuli}); these visualizations predominantly belong to the category of static infographics, with varying levels of text annotation. The dataset was compiled by scraping various real-world visualization sources online, including government reports, infographic blogs, news media, and scientific journals. The diversity and distribution of these visualizations offer a comprehensive representation of data visualizations ``in the wild'', with a significant portion sourced from infographics or social media, making them more accessible to non-expert users. Arunkumar et al. extensively pre-coded these visualizations with labels indicating features such as human-recognizable objects, data-ink ratio, underlying data structures, and visual encodings, aligned with the visual taxonomy developed by Borkin et al.~\cite{borkin2013makes,borkin2015beyond} for MASSVIS.$^1$

\subsection{Set-up and Participants} 

As discussed above, we utilized 100 visualizations curated by Arunkumar et al.~\cite{10294209}, resizing them while maintaining aspect ratios to a maximum dimension of 1000 pixels. Participants completed the study in three phases (see Fig.~\ref{fig:survey}\footnote{\label{footno1}See Supplemental Material for (i) visual taxonomy, (ii) full survey question set, (iii) demographics, (iv) break periods, (v) data coding details, (vi) statistical modelling and mediation analysis.}). 

\anj{
In Phase 1, participants must complete 50 trials (randomly selected visualizations) in succession that determine their at-a-glance comprehension of the visualization, by examining whether they are able to accurately identify the base chart type, data types, and the creation source (e.g.: government, news media, etc.). This helps establish a baseline for understanding how effectively they process visual information in the context of the study. In Phase 2, participants complete 50 trials using the same subset of 50 stimuli seen in Phase 1, presented in a random order. First, they view a stimulus for 45 seconds, and self-report the occurrence of different types of mind-wandering via key-press (see Sec.~\ref{sec:dc}). Since mind wandering is considered to be a dynamic measure that can potentially reflect a viewer’s engagement patterns, participants rate their overall cognitive and emotional engagement levels. Cognitive and affective engagement can potentially be considered mechanisms which shape users’ overall  impressions of their emotional response and their perceptions of design-quality~\cite{arunkumar2023image}; hence these ratings are collected as well. To this, we added the dimensions of trust, as they pose a critical factor in user engagement with visualizations; if users do not trust the information presented, their engagement is likely to diminish, regardless of the visualization's design quality or their emotional response~\cite{pandey2023you}. Finally, we additionally ask participants to verbalize their cogntive and affective visualization impressions using adjectives and a short-term recall task, as the initial set of tasks are numerical ratings based. In Phase 3, one week after the study, participants complete a visual recognition task to test long-term recall. In this manner, we test mind wandering influence on different modalities of post viewing measures collected.
} Each participant underwent three training trials, followed by the main study. They viewed 50 random stimuli from the dataset in Phases 1 and 2, including two attention checks, and were presented with the full dataset in Phase 3.

We also conducted a pilot study with 3 participants to validate the study design. For the main study, we recruited 106 graduate students (26.4 $\pm$ 3.9 y/o) from Arizona State University, enrolled in the data visualization course. All participants had normal color vision. Participants were given the option to earn extra credit equivalent to 1\% of their total grade by completing all phases of the survey within the initial three weeks of the course commencement. To ensure participation was fully autonomous and non-compulsory~\cite{shen2021student}, students could alternatively choose from several extra credit opportunities, including non-research options. Study duration averaged roughly 10:09 ($\pm$2:18) minutes for Phase 1, 93:20 ($\pm$6:38) minutes for Phase 2, and 3:27 ($\pm$0:52) minutes for Phase 3. Participants were also allowed up to two 15-minute breaks during Phase 2, to prevent study fatigue. Each visualization was viewed at least 50 times.$^1$

\subsection{Data Coding}
\label{sec:dc}

\textbf{Visualization Features:} We utilized the taxonomy data from the dataset$^1$ as independent variables for analysis. All label values were mapped to either binary (0/1) or ternary (0/1/2) scales before constructing the model.$^1$

\noindent \textbf{At-A-Glance Comprehension:} Arunkumar et al.~\cite{10294209} found that visualizations may be internalized differently, either as images or information representations, based on various visual features. To control for this, we measured at-a-glance understandability. Using metadata from the dataset taxonomy, including chart type, data type, and creation source, we evaluated participant answers as correct (1) or incorrect (0). For partially correct answers, we assigned a value of 0.5. These attributes were also treated as independent variables during analysis.

\noindent \textbf{Mind Wandering:} \anj{In a pre-study pilot, we asked participants to report only relevant vs. irrelevant mind wandering occurrence. At the end of that pilot study, participants also reported what they considered ``relevant" mind wandering, in context how it aligned with their visualization viewing goals. We found that their reports (see Figure~\ref{fig:teaser} for an example) could be broadly linked to (1) chart appearance: anything that discusses the aesthetic impact of the visualization, (2) chart data: anything pertaining to data-oriented or cognitive tasks such as trend identification, and (3) chart topic: anything pertaining to external knowledge integration or affective engagement that is related to the overall concept of the chart. Therefore, we asked participants to explicitly report the type of relevant mind wandering as well as (4) irrelevant mind wandering in the main study. This was done as self-reports via key presses during the 45 second observation period. In addition to frequency,} temporal variables denoting the earliest instance of reported mind wandering were computed as: \textit{total observation time} $-$ \textit{earliest-report timestamp}. These variables were treated as separate mediating variables during modeling.

We discuss the dependent variables (collected measures) below:

\noindent \textbf{Trust:} Utilizing a 5-item measure~\cite{pandey2023you} validated over the MASSVIS dataset~\cite{borkin2013makes}, we collected post-viewing ratings on credibility, clarity, reliability, familiarity, and confidence using a 5-point Likert scale.

\noindent \textbf{Engagement:} Post-viewing cognitive~\cite{mahyar2015towards} and affective~\cite{lee2022affective} engagement were measured using 5-point Likert scales based on prior literature, providing a direct summary variable for comparison with mind-wandering self-reports.

\noindent \textbf{Design Quality:} We evaluated design quality using a survey instrument adapted from Moshagen et al.~\cite{moshagen2010facets}, assessing simplicity, diversity, and craftsmanship with 5-point Likert data corresponding to the highest factor loadings under each dimension.

\noindent \textbf{Emotional Response:} Emotional engagement was measured using SPANE (The Scale of Positive and Negative Experience)~\cite{rahm2017measuring}, assessing post-viewing positive and negative affect on a 5-point Likert scale.

\noindent \textbf{Adjective Coding:} Participants described visualizations based on aesthetic appeal and communicative utility using three adjectives each. This qualitative approach aimed to identify combinations of visualization features that could restore attention, countering the effects of mind wandering on both cognitive and affective engagement. Adjectives were coded based on strength and valence (see Table~\ref{tab:2}). Drawing on prior work in Sentiment Analysis~\cite{ruppenhofer2014comparing,ruppenhofer2015ordering}, we grouped adjectives into intensity scales based on positive/negative polarity. An intensity scale for the adjectives collected in our study was established using semantic similarity-based labeling approaches~\cite{sheinman2013large,siddique2019computing}. Three expert annotators (Fleiss' $\kappa$ \anj{denoting annotator agreement} = 0.79, ~\cite{kilicc2015kappa}) further validated the adjective coding. Adjectives were represented by 2D vectors <valence, strength>, where valence [-1,1] and strength [1,6] were continuous quantitative dimensions.

\noindent \textbf{Short-Term Recall Coding:} Participants provided single-sentence responses about the visualization's main takeaway and the most salient feature. Three annotators (Fleiss' $\kappa$ = 0.72) coded responses for design elements (codes 1-8 from Table~\ref{tab:4}) and personal opinions (9: like/dislike; 10: perceived trends; 11: external knowledge). This coding reflected which visualization features were most robust to mind-wandering. Responses were represented as 11-dimensional vectors, with each dimension assigned a value of 0/1 based on mentioned codes.

\noindent \textbf{Long-Term Recall:} Participants performed a visual recognition test , one week after the study (similar to the methodology followed by Borkin et al.~\cite{borkin2015beyond}). \anj{This assessed participants' ability to identify previously encountered visual stimuli by recalling from long-term memory.} Accuracy Values: \textit{<TP (1) / FP (2) / TN (3) / FN (4)>}. 

\section{Results \& Discussion}
\label{sec:5}

In this section, we present results for research questions \textbf{\textit{RQ1}} –- \textbf{\textit{RQ3}}, assessed based on hypotheses outlined in Section \ref{sec:3}. Using R (4.3.3), we conducted several iterations of multivariate SEM (Structural Equation Modeling)~\cite{joreskog1973analysis} with Robust Maximum Likelihood Estimate, while maintaining a statistical power of 0.8\anj{~\cite{hair2011pls}}. Below, we report results for models demonstrating the best fit to the data, determined by established fit indices and statistical significance.$^1$ Initially, we developed latent variable constructs for visualization design elements and at-a-glance comprehension scores based on SEM factor loadings, as detailed in Table \ref{tab:4}. Each latent variable was then used as a stand-alone predictor. We refrained from constructing latent variables for dependent variables due to low goodness of fit index. However, for mind-wandering self-reports (tested as a mediator), we analyzed:
(i) Frequency of each mind-wandering type,
(ii) Summative aggregation of 'relevant' mind-wandering frequency,
(iii) Overall mind-wandering frequency, and
(iv) Temporal distribution of mind wandering.

\begin{table}[!h]
    \centering
    \scriptsize
    \caption{\anj{Latent Independent Variable constructs used in SEM.}}
    \begin{tabular}{p{0.2\columnwidth}p{0.5\columnwidth}}
    \toprule
        \textbf{Latent Variable Construct} & \textbf{Constituent Variables} \\
        \midrule
        1. Encodings & complex glyphs, human recognizable objects, human depiction\\
        2. Realism & photo-realism, 3D, skeuomorphism \\
        3. Annotation & arrows, shading, other highlights\\
        4. Text & title, caption, key, labels, text volume\\ 
        5. Emphasis & data redundancy, message redundancy, data source, data ink ratio\\
        6. Structure & axes, gridlines, background color, number of colors\\
        7. Complexity & density, data volume, dimensionality\\
        8. Comprehension & accuracy in identifying chart-type, data-type, creation source at-a-glance\\      
        \bottomrule        
    \end{tabular}   
    \label{tab:4}
        \vspace{-2em}
\end{table}

\subsection{RQ1: Mind Wandering and User Experience}

\noindent \textbf{\textit{RQ1 asks: does mind wandering during visualization exploration influence viewer experience?}}

\noindent \textbf{Summary of Findings: }Overall, mind wandering significantly impacts all collected post-viewing measures.\footnote{\anj{All study results reported here are significant, $p<0.01$ or $p<0.05$.}} Higher frequencies of mind wandering and earlier reported instances are associated with stronger effects. Notably, the accuracy of short-term recall and trust dimensions (credibility, reliability, and confidence) show the strongest negative impact regardless of mind-wandering type. `Relevant' mind wandering, especially pertaining to the chart's topic, moderately enhances affective engagement and weakly affects emotional response, aesthetic appeal adjectives, and accuracy of long-term recall. However, reported cognitive engagement declines. Mind wandering also moderately negatively affects design quality ratings and communicative utility adjectives.

\noindent \textbf{Details of Analysis: } To examine this question more closely, we focus on normalized regression results from SEM, between mind-wandering frequency/earliest instance and collected measures. 

\noindent \textbf{H1a: } \textit{Higher levels of mind wandering will correspond with lower ratings of trust, engagement, emotional response, and design-quality.
}

\begin{figure}[!ht]
    \includegraphics[width=0.4\textwidth]{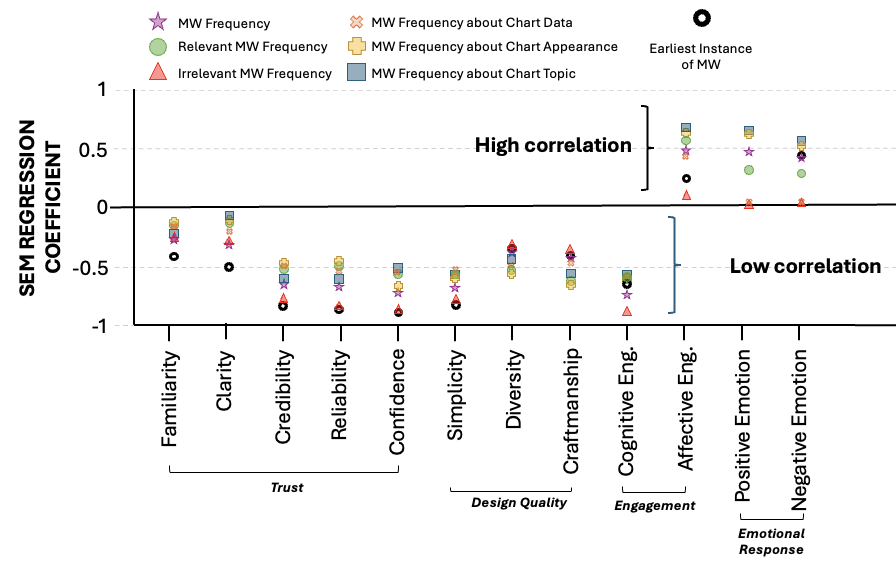}
    \caption{Scatterplot denoting the Regression Coefficient ($\beta$) results (y-axis) for observed/latent variables for mind wandering (MW) vs. collected quantitative post-viewing measures (x-axis). Color/shape denotes the mind wandering variable considered. We observe that all the `relevant mind-wandering' variables cluster close together, as do irrelevant mind-wandering frequency, earliest occurrence, and aggregated frequency.}
    \label{fig:RQ1a}
        \vspace{-1em}
\end{figure}

\noindent \textbf{Trust: }We found that the trust dimensions of credibility ($\beta = -0.65$, $f^2 = 0.23$), reliability  ($\beta = -0.69$, $f^2 = 0.33$), and confidence  ($\beta = -0.73$, $f^2 = 0.41$) show strong negative associations with respect to mind wandering, as shown in Fig.~\ref{fig:RQ1a}. The effect strength decreased when mind wandering was pertinent to chart data ($\overline\triangle\beta = +0.14$) and increased for irrelevant mind wandering ($\overline\triangle\beta = -0.08$). \anj{For familiarity and clarity, however, we found weak negative effects ($\overline\beta = -0.34$, $p=0.043$), irrespective of the type of mind wandering;} this may be due to the reliance of these dimensions on prior knowledge and comprehension rather than immediate attentional focus. We also found an increase in effect strength across all trust dimensions, when the first reported instance of mind wandering occurred at an earlier time stamp while viewing ($\overline\triangle\beta = -0.12$, $\overline\triangle f^2 = +0.06$).

\noindent \textbf{Engagement: }Mind wandering was found to have mixed effects on engagement \anj{($p=0.037$)}. Lower levels of cognitive engagement were strongly/moderately associated with higher frequency ($\beta = -0.73$, $f^2 = 0.39$) and earlier occurrence ($\beta = -0.62$, $f^2 = 0.24$) of mind wandering, respectively, as shown in Fig.~\ref{fig:RQ1a}. Participants reported feeling like they had been ``exposed to the topic'' or ``trying to understand the encodings'' of the chart when a higher frequency or earlier first report of mind-wandering occurred; irrelevant mind wandering exacerbated this ($\overline\triangle\beta = -0.15$), while that pertinent to chart data begat improvement ($\overline\triangle\beta = +0.14$). In general, participants did not report very high levels of cognitive engagement (i.e., ``taking a decision" based on chart content); this could have arisen from the non-assignation of a specific cognitive task to perform. However, in the case of affective engagement, moderate positive effects were seen for mind wandering pertinent to the chart's appearance ($\beta = +0.59$, $f^2 = 0.26$) or topic ($\beta = +0.62$, $f^2 = 0.30$), with participants feeling like their beliefs about the chart topic had been ``strengthened" or ``juxtaposed." Participants reported moderate--high affective engagement across the charts overall, potentially due to significant variation in chart topics, themes, and styles presented.

\noindent \textbf{Emotional Response: }Both positive and negative emotional responses had moderate positive associations with mind wandering pertinent to a chart's topic (positive: $\beta = +0.62$, $f^2 = 0.21$; negative: $\beta = +0.55$, $f^2 = 0.18$) and aesthetic appearance (positive: $\beta = +0.59$, $f^2 = 0.22$; negative: $\beta = +0.51$, $f^2 = 0.16$), respectively. This is attributed to the association of emotion with affective engagement, which showed a similar relationship with mind wandering. Irrelevant/pertinent to data mind wandering had negligible effects on emotional response in general. Overall, mind wandering frequency was found to have a weak positive association with positive/ negative ($\overline\beta = +0.45$, $f^2 = 0.06$) emotions, as shown in Fig.~\ref{fig:RQ1a}. The earliest occurrence of mind wandering showed comparable effects to frequency.

\noindent \textbf{Design Quality: }We found that mind wandering frequency only had a weak negative influence on diversity and craftsmanship ($\overline\beta = -0.35$), comparable with regression based on the earliest occurrence. Mind wandering pertinent to the chart's appearance increased the effect strength ($\overline\triangle\beta = -0.12$, $\overline\triangle f^2 = +0.07$), while the other types had a comparable impact, as shown in Fig.~\ref{fig:RQ1a}. This might have diverted attention from evaluating the diversity and craftsmanship of the content, potentially leading to a more negative perception of these dimensions. Simplicity ratings, however, were moderately negatively affected by an increased frequency ($\beta = -0.61$, $f^2 = 0.21$), and exacerbated by earlier occurrences of mind wandering ($\overline\triangle\beta = -0.13$, $\overline\triangle f^2 = +0.05$), as well as when mind wandering was irrelevant ($\overline\triangle\beta = -0.11$, $\overline\triangle f^2 = +0.09$), indicating a disruption to the individual's ability to process and understand the information presented. 

\textbf{\textit{Hence, this hypothesis is supported for trust and design-quality, partially for engagement, and not supported for emotional response.}}

\noindent \textbf{H1b: } \textit{Participants experiencing more frequent mind wandering will describe the communicative utility of stimuli with strong negative adjectives, and aesthetic appeal with strong positive adjectives.}

We noticed that in every trial, participants consistently maintained the relative positivity or negativity (valence) and the intensity (strength) of adjectives, with regards to aesthetic appeal and communicative effectiveness. In total, participants generate 521 unique adjectives for aesthetic appeal, and 582 unique adjectives for communicative utility.$^1$ During training, participants received an extensive list of sample adjectives to understand the type of descriptions required. Incorrectly assigned adjectives were found in less than 5\% of trials, with no instances of all adjectives being inaccurately targeted. Therefore, all trials were included in the analyses. Instances of incorrect adjective assignments were replaced with duplicate entries of correct adjective inputs from the participant within the respective trial. 

\begin{figure}[!ht]
    \includegraphics[width=0.4\textwidth]{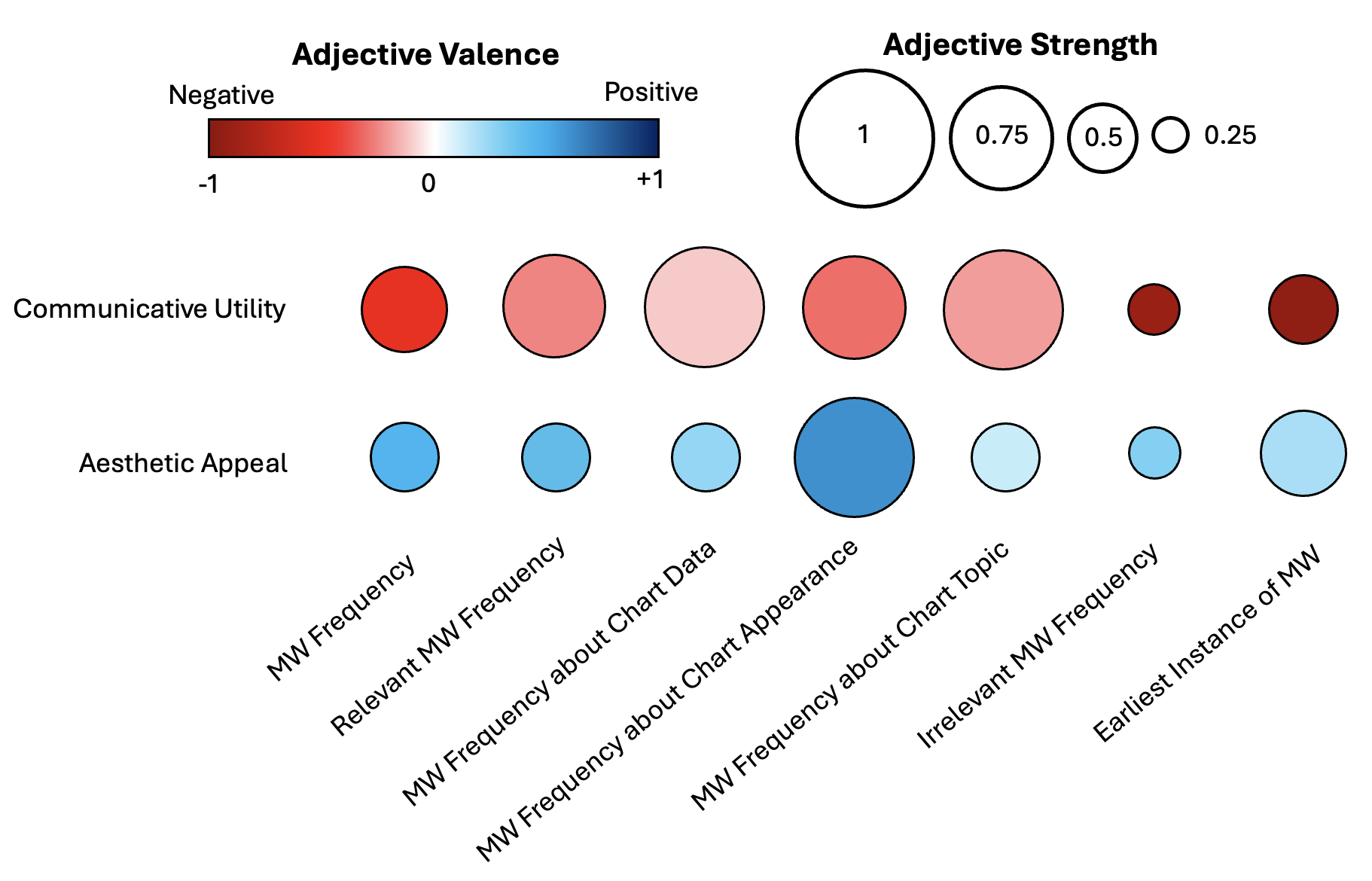}
    \caption{Correlogram denoting how mind wandering (MW) affects the normalized valence (color) and strength (size) of adjective descriptions. We note that irrelevant MW causes the greatest decrease in adjective strength. MW on chart appearance increases positive valence of aesthetic appeal, irrelevant MW and earliest occurrence of MW increases negative valence of communicative utility. MW on chart data/topic has the weakest effects overall. }
    \label{fig:RQ1c}
        \vspace{-1em}
\end{figure}

\begin{figure*}[!ht]
\centering
    \includegraphics[width=0.7\textwidth]{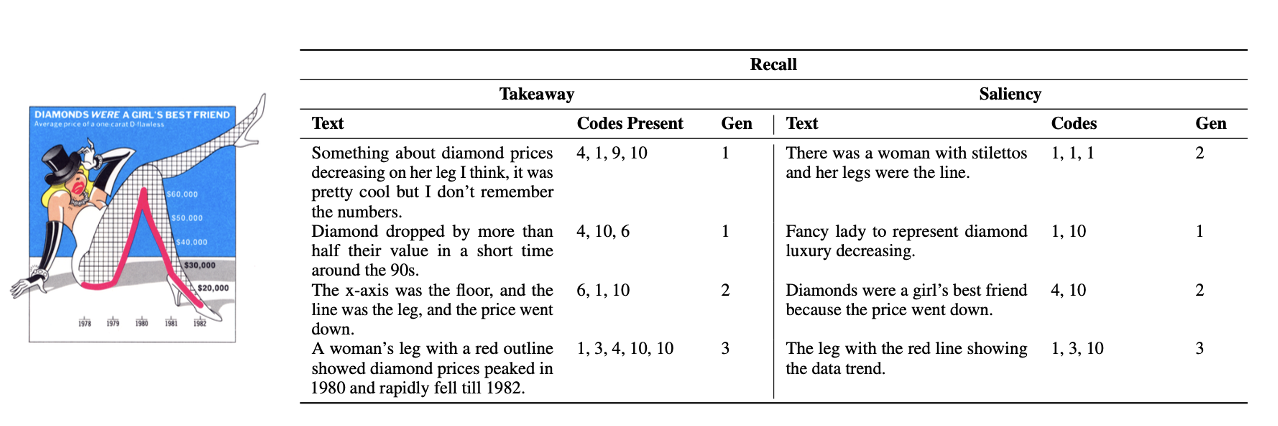}
    \vspace{-1em}
    \caption{Example of how short-term recall is coded for visualization takeaways and most salient visualization elements. (Gen: Generality Score).}
    \label{fig:RQ1d1}
        \vspace{-2em}
\end{figure*}

We found that when mind wandering frequency increased, adjectives for communicative utility had a more negative valence ($\beta = -0.58$, $f^2 = 0.20$; $\overline{valence} = -0.68$), though adjective intensity could vary between weak--high ($\overline{strength} = 2.54$); if irrelevant mind wandering instances were more prevalent ($\beta = -0.64$, $f^2 = 0.23$), the effect was increased ($\overline{valence} = -0.77$), with an increased usage of moderate--high intensity adjectives ($\overline{strength} = 3.96$) (see Fig.~\ref{fig:RQ1c}). The presence of irrelevant mind wandering within the first 15 seconds of viewing further bolsters this effect ($\beta = -0.71$, $f^2 = 0.32$; $\overline{valence} = -0.81$; $\overline{strength} = 4.32$). Participants might struggle to engage with the visualization's underlying data during mind wandering, potentially leading to a stronger negative valence assigned to communication.

Adjectives for aesthetic appeal had a more positive valence ($\beta = 0.43$, $f^2 = 0.08$; $\overline{valence}=0.32$), for increased mind wandering frequencies, though adjective intensity was mostly weak ($\overline{strength}=1.63$). Irrelevant mind wandering instances maintained an overall positive valence($\overline{\triangle{valence}}=-0.11$), but decreased adjective strength further ($\overline{\triangle{strength}}=-0.37$), as shown in Fig.~\ref{fig:RQ1c}. However, mind wandering pertinent to a chart's appearance significantly increased the average valence ($\overline{\triangle{valence}}=+0.22$) and strength ($\overline{\triangle{strength}}=+0.64$) of aesthetic appeal adjectives. We posit that examination of the chart's appearance has led to a heightened appreciation for aesthetic aspects, increasing the strength and valence of positive responses. 

\textbf{\textit{Hence this hypothesis is partially supported over both communicative utility and aesthetic appeal adjectives.}}


\begin{figure}[!ht]
    \includegraphics[width=0.4\textwidth]{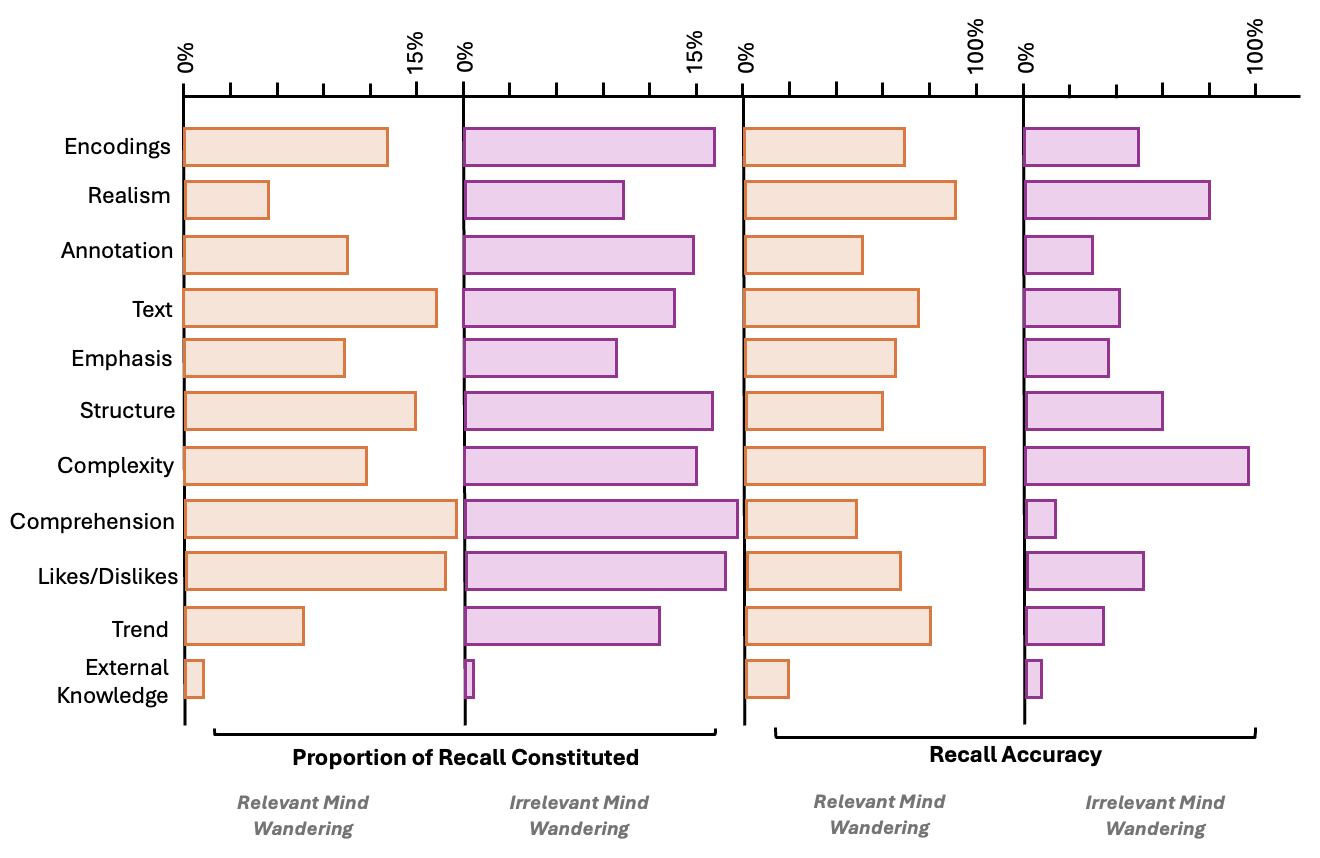}
    \caption{Small multiple bar charts for both relevant (purple) and irrelevant (orange) mind wandering frequencies. These illustrate how mind wandering influences the: (i) proportion of codes occurring in recall, and (ii) if the coded recall content is accurate. Text, Emphasis, and Trends show the largest decrease in accuracy for irrelevant mind wandering, while Complexity and Encodings increase in proportion the most.}
    \label{fig:RQ1d2}
    \vspace{-0.5em}
\end{figure}

\noindent \textbf{H1c: } \textit{Participants experiencing more frequent mind wandering will exhibit poorer short-term and long-term recall accuracy, as well as decreased specificity of short-term recall.}

\anj{As mentioned in Section~\ref{sec:dc}, short-term recall was coded into 11D vectors based on mentions of design elements or personal opinions in participant responses. To quantify recall accuracy, annotators evaluated the dimensions marked as `present' (value=1) against the visualization content, assigning -1 for inaccuracies. Overall recall accuracy was computed by averaging these values. Recall generality was coded as follows: 1 for mentions of a design element's presence, 2 for mentions of a design element's appearance, and 3 for mentions of a specific data point referenced by a design element. The recall generality was averaged across all ‘present’ dimensions (see Fig.~\ref{fig:RQ1d1}).}


Visualization takeaways were 15.60 words long, with 3.12 codes mentioned, on average. Figure~\ref{fig:RQ1d2} summarizes the strong negative impact of increased mind wandering frequency on takeaway accuracy ($\beta = 0.43$, $f^2 = 0.08$) and takeaway generality ($\beta = 0.43$, $f^2 = 0.08$), over different code dimensions. We noted that in general, higher irrelevant mind wandering frequency increased the mention of encoding elements (8.2\%$\rightarrow$10.7\%) and complexity (3.4\%$\rightarrow$7.1\%) of the chart, while decreasing the mention of emphasis elements (6.3\%$\rightarrow$4.8\%) and perceived trends(4.9\%$\rightarrow$4.5\%). Text was frequently mentioned in takeaways (10.3\%) when present in the visualization, however, the accuracy ($\overline{\downarrow{21.35\%}}$) and generality ($\overline{\downarrow{27.82\%}}$) of text, emphasis elements, and perceived trends decreases with increased frequency of mind wandering. \anj{The temporal distribution of mind wandering did not significantly affect takeaways. Mind wandering may disrupt cognitive processes needed to synthesize information from visualizations, leading to fragmented or superficial interpretations that are more likely to be inaccurate.}

\anj{Reports of the most salient visual element showed that all types of mind wandering had comparable, weak negative impacts on accuracy ($\beta = 0.38$, $f^2 = 0.04$), but no significant impact on generality or the temporal distribution of mind wandering. Encoding, realism, and annotation made up 87\% of recall content, reflecting the focus on visualization appearance. Overall, saliency was more robust to mind wandering than takeaways.}

\anj{We also found that mind wandering had a non-significant impact on long-term recall performance. Specifically, relevant mind wandering had a weakly positive impact on long-term recall accuracy ($\beta = 0.39$, $f^2 = 0.06$), while irrelevant mind wandering showed no significant effect. Task-related mind wandering may act as elaborative rehearsal, facilitating deeper encoding and enhancing memory consolidation. In contrast, irrelevant mind wandering lacks cognitive processing related to the task at hand and thus does not significantly affect memory performance.}


\textbf{\textit{Hence this hypothesis is partially supported for short-term recall, and not supported for long-term recall.}}

\subsection{RQ2: Design Elements and Nature of Mind Wandering}

\noindent \textbf{\textit{RQ2 asks: how do design elements influence the type, frequency and temporal distribution of mind-wandering?
}}

\noindent \textbf{Summary of Findings: } Aesthetic novelty, like complex glyphs and human-recognizable objects, significantly increased mind-wandering frequency. Interestingly, the text's impact wasn't linear; both very low and high volumes correlated with more mind wandering, while moderate volumes showed less. Axes, data, and message redundancy had the strongest negative effects. Irrelevant mind wandering was common in dense visualizations with high data volumes. Relevant mind wandering occurred in low data-ink ratio charts. Temporal distribution was influenced by text volume and data ink ratio, with lower values triggering earlier mind wandering instances.

\noindent \textbf{Details of Analysis: } To examine this question more closely, we focus on the normalized regression results from our SEM model, between design elements and mind-wandering frequency/time. 

\begin{figure}[!ht]
\centering
    \includegraphics[width=0.37\textwidth]{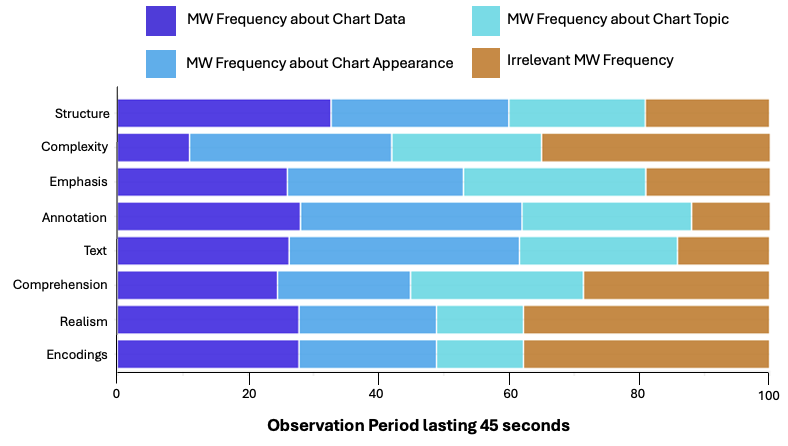}
    \caption{Stacked bar chart representing relative occurrence of different types of mind wandering (MW), in the presence of different design elements. Overall, chart appearance is most prevalent; however, the occurrence of irrelevant mind-wandering is very high for Encodings and Photorealism, which hinder at-a-glance comprehension.}
    \label{fig:RQ2a}
        \vspace{-2em}
\end{figure}

\noindent \textbf{H2a: } \textit{\anj{Visualizations that prominently feature design elements like complex Encodings and Realism but receive low ratings for At-A-Glance Comprehension are more likely to induce irrelevant mind wandering earlier in the observation period.}}

\begin{figure}[!ht]
\centering
    \includegraphics[width=0.35\textwidth]{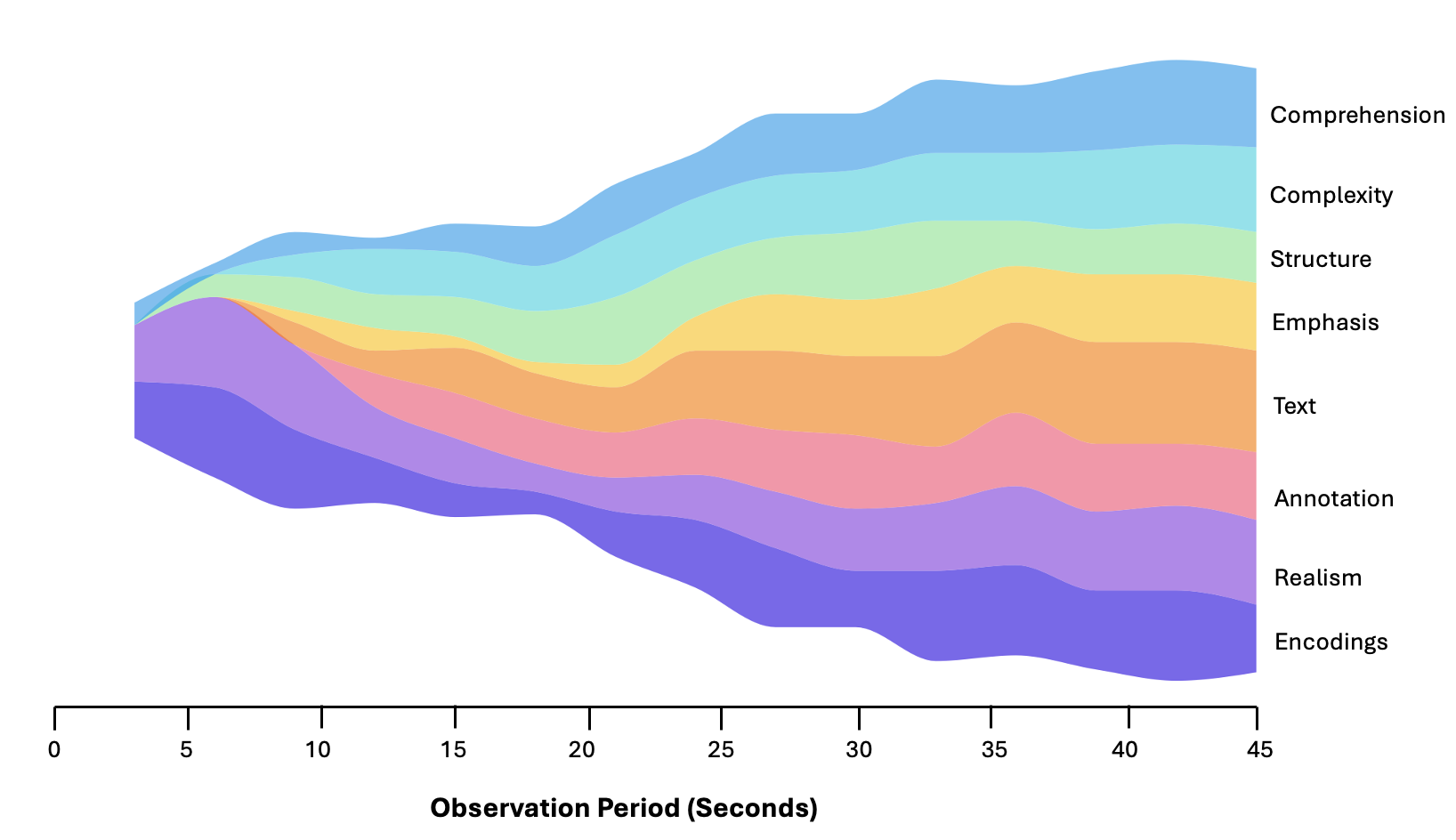}
    \caption{Streamgraph illustrates the temporal distribution of mind wandering instances throughout the observation period, with different design elements affecting the probability of reporting mind wandering over time. While the presence of encoding and realism elements increases the likelihood of mind wandering at the beginning, text elements are more likely to result in late-stage mind wandering.}
    \label{fig:RQ2b}
        \vspace{-1em}
\end{figure}

The presence of aesthetic novelty elements (represented by the variables Encodings and Realism) strongly increased the frequency of irrelevant mind-wandering ($\uparrow$34.23\%, $\beta = 0.68$, $f^2 = 0.38$), as shown in Figure~\ref{fig:RQ2a}. There was also a moderate increase in the frequency of mind wandering pertinent to a chart's appearance ($\uparrow$18.97\%). Additionally, we observed that initial mind-wandering reports occurred closer to the beginning of the observation period ($\overline{t} = 6.78s$, $\beta = 0.56$, $f^2 = 0.23$), particularly for irrelevant mind-wandering. On further analyzing the temporal distributions of different types of mind-wandering (see Figure~\ref{fig:RQ2b}), we saw that there is an uptick in the total number of mind-wandering instances for the last 10 seconds of observation ($\uparrow$14.33\%), particularly concerning chart appearance ($\uparrow$22.19\%). A potential explanation could be that the presence of such elements may evoke a sense of curiosity or fascination, leading individuals to engage in exploratory or imaginative thought processes, causing them to become more prone to experiencing mind wandering episodes.

Poor at-a-glance comprehension was found to have a strong association with increased mind-wandering frequency ($\uparrow$26.37\%, $\beta = -0.71$, $f^2 = 0.39$). The effects across both relevant and irrelevant mind wandering were comparable. Additionally, we observed that initial mind-wandering reports occurred closer to the beginning of the observation period ($\overline{t} = 6.31s$, $\beta = 0.66$, $f^2 = 0.25$) for all types of mind wandering. A potential explanation is that individuals may experience cognitive strain and frustration when struggling to grasp the meaning of presented information, triggering a negative affective response and leading to mind wandering as they disengage from interpretation.

\textbf{\textit{Hence this hypothesis is supported.}}

\noindent \textbf{H2b: } \textit{\anj{Visualizations that incorporate design elements of Structure, Annotation, Emphasis, and Realism are less likely to induce irrelevant mind-wandering. If mind-wandering does occur, it may typically manifest towards the end of the observation period.}}

\anj{Visualizations resembling their abstract canonical impressions (internal reference images representing chart types) showed moderate resistance to mind wandering overall ($\downarrow$19.76\%, $\beta = -0.53$, $f^2 = 0.20$). However, no significant effects were found on the temporal distribution of mind wandering. Design element features may demand more attention compared to structural features mapping visualizations to canonical abstractions, influencing attentional resource allocation differently but not impacting temporal mind wandering dynamics.}

\anj{The presence of annotation elements significantly decreased the frequency of irrelevant mind-wandering ($\downarrow$40.66\%, $\beta = -0.73$, $f^2 = 0.41$). In a secondary model focusing on text annotation alone, both extremely low and high text volumes resulted in significant increases in irrelevant ($\uparrow$9.85\%) and relevant mind-wandering ($\uparrow$6.49\%), respectively. Regarding temporal distributions, irrelevant mind-wandering occurred later in the observation period ($\overline{t} = 28.78s$), especially in the text-annotation only model ($\overline{t} = 33.63s$). We hypothesize that text presence increases familiarity with content, potentially reducing cognitive vigilance and increasing mind-wandering susceptibility, especially in later stages. Conversely, low text conditions might hinder chart comprehension, leading to mind wandering.}


We observed that the presence of emphasis elements strongly decreases the frequency of all types of mind wandering, in a comparable manner ($\downarrow$34.31\%, $\beta = -0.76$, $f^2 = 0.37$). We also noted that mind-wandering first occurs only within the last 20 seconds of the observation period ($\overline{t} = 31.07s$, $\beta = -0.63$, $f^2 = 0.28$), as shown in Fig.~\ref{fig:RQ2b}. Emphasis elements, therefore, promote sustained attention and task engagement, leading to decreased mind wandering episodes. 

\textbf{\textit{Hence this hypothesis is partially supported for Structure, and fully supported for the other constructs.}}

\noindent \textbf{H2c: } \textit{\anj{Visualizations characterized by factors such as high data volume, text volume, visual density, and a lower data-ink ratio are more likely to induce irrelevant mind wandering early in the observation period.
}}
We found that visualizations with higher associated complexity were moderately susceptible to mind wandering ($\uparrow$15.88\%, $\beta = 0.48$, $f^2 = 0.17$); surprisingly, we found that the effects were more pronounced for relevant($\uparrow$19.31\%) rather than irrelevant ($\uparrow$14.27\%) mind-wandering. A similar pattern was observed with regard to the temporal distributions-- relevant ($\overline{t} = 7.34s$, $\beta = 0.53$, $f^2 = 0.21$) mind wandering was seen to occur at earlier timestamps, in comparison to irrelevant ($\overline{t} = 10.31s$, $\beta = 0.47$, $f^2 = 0.18$) mind wandering (though both occurred within the first 15 seconds of the observation period). We suggest that relevant mind wandering may be particularly pronounced here because individuals are actively engaged in trying to comprehend and make sense of the complex information presented (leading to more frequent episodes of task-related distraction), whereas irrelevant mind wandering may occur later as attentional resources become depleted. \textbf{\textit{Hence this hypothesis is not supported.}}

\subsection{RQ3: Mind Wandering as a Mediator}

\noindent \textbf{\textit{RQ3 asks: does mind wandering mediate the relationship between design elements and post-viewing measures?}}

\noindent \textbf{Summary of Findings: } Overall, we found that the frequency and earliest reported instance of mind-wandering acts as a partial mediator between a majority of the direct effects seen in our model, i.e., between the design elements and post-viewing measures collected.

\begin{figure}
\centering
    \includegraphics[width=0.35\textwidth]{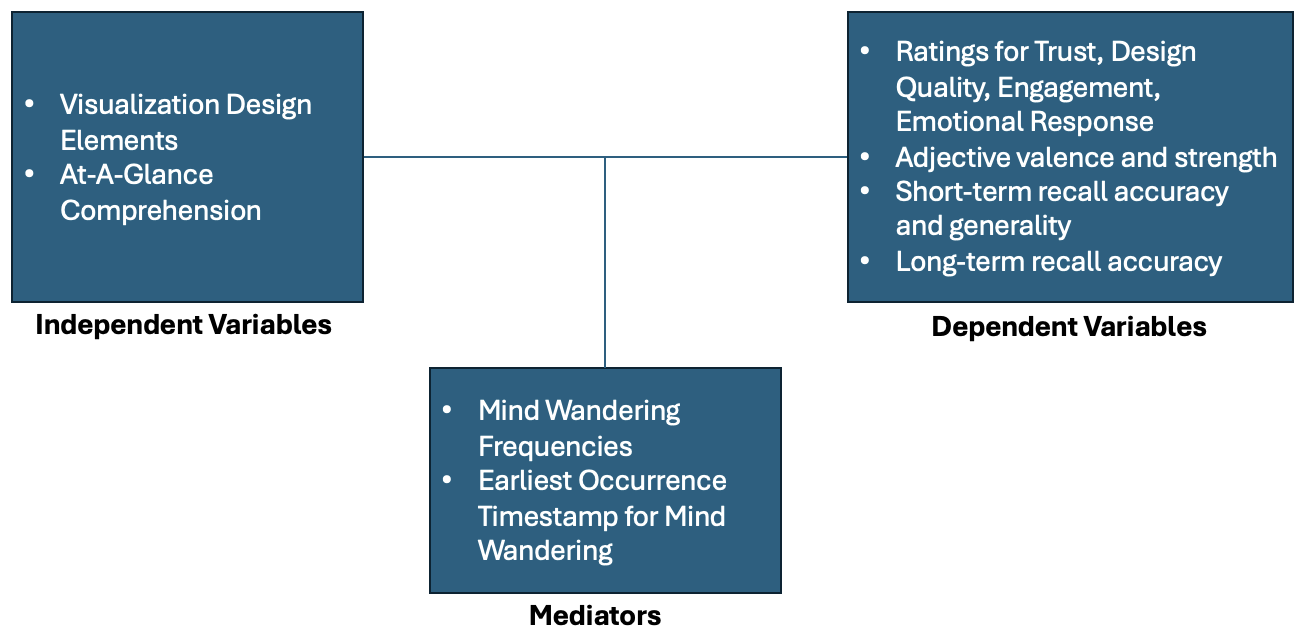}
    \caption{SEM analysis structure.}
    \label{fig:RQ3a}
        \vspace{-2em}
\end{figure}

\begin{figure}[!ht]
\centering
    \includegraphics[width=0.4\textwidth]{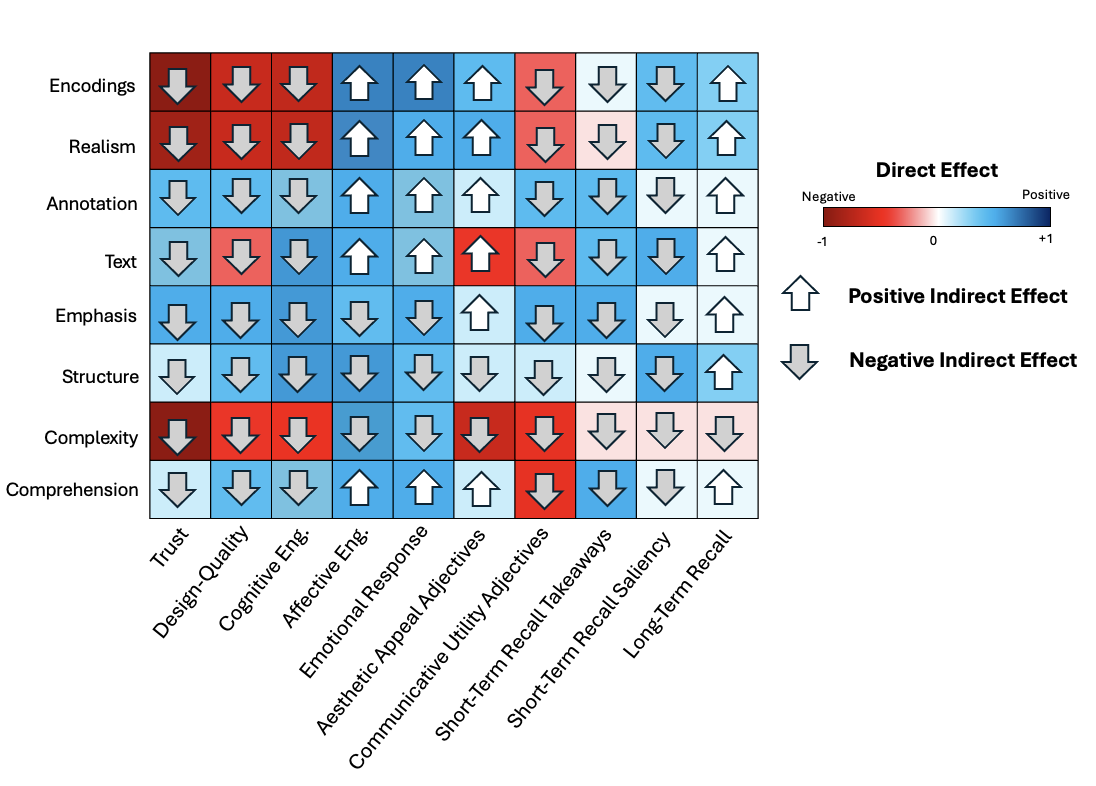}
    \vspace{-1.1em}
    \caption{Comparison of Direct and Indirect Mediating Effects. Heatmap cells represent the strength and directionality of direct effect. Mind wandering acts as a partial mediator across all relationships between the independent variables (design elements) x dependent variables (post-viewing measures). The directionality of the direct effect is not reversed by the indirect effect; rather, its intensity is strengthened/weakened to varying degrees. We represent the directionality of the indirect effect with arrow overlays ($\uparrow$: same / $\downarrow$: opposite direction as the direct effect).}
    \label{fig:RQ3b}
        \vspace{-1em}
\end{figure}

\noindent \textbf{Details of Analysis: } To examine this question more closely, we focus on the normalized regression results from our SEM model. We compare the magnitude of regression coefficients for the direct path (independent--dependent variable) and indirect path (independent--mediator--dependent variable). We further perform bootstrapping to validate the significance of the indirect effects. Partial mediation is achieved when the direct and indirect effects are significant, while full mediation is achieved when only the indirect effect is significant~\cite{mackinnon2007mediation,baron1986moderator}. We illustrate the overall structure of our SEM in Figure~\ref{fig:RQ3a}. We specify model paths connecting every possible combination of <independent--mediator--dependent> path in our model. Hence, to succinctly summarize model results, below, we discuss the magnitude and directionality of direct/indirect effects observed (also see Fig.~\ref{fig:RQ3b}), and report the averaged coefficient values across each set of paths discussed.$^1$

\noindent \textbf{H3: } \textit{Mind wandering will partially negatively mediate user ratings of engagement, emotional response, design-quality and trust, communicative utility and aesthetic appeal adjectives, and the accuracy of short-term and long-term recall.}

\noindent \textbf{Engagement and Emotional Reponse: }Encodings, Realism, and Complexity had direct negative effects on cognitive engagement ($\overline{\beta} = -0.68\pm0.07$), but had direct positive effects for affective engagement and emotional response ($\overline{\beta} = 0.64\pm0.09$). Additionally, Comprehension and Annotation showed moderate positive effects for all three variables ($\overline{\beta} = 0.48$). Text, Engagement, and Structure strongly boost cognitive engagement ($\overline{\beta} = 0.67\pm0.11$), but showed weaker positive effects relating to affective engagement and emotional response ($\overline{\beta} = 0.52$). The temporal distribution of mind wandering is seen to partially mediate paths for affective engagement and emotional response positively ($\overline{\beta} = 0.48$), and cognitive engagement negatively ($\overline{\beta} = -0.53$). The positive mediation of affective engagement by mind wandering may seem counterintuitive at first, but mind wandering may prompt users to reflect on the emotional significance of the visualization's content. This introspective processing can evoke emotional responses such as empathy, nostalgia, or aesthetic appreciation, which enhance affective engagement. The frequency of mind wandering negatively partially mediates all the aforementioned paths ($\overline{\beta} = -0.49\pm0.10$). Occasional mind wandering during less critical moments of interaction may not significantly impact affective engagement or emotional response, but frequent mind wandering during emotionally salient content may disrupt these dimensions. \anj{The indirect effects of both the reporting frequency and temporal distribution of mind-wandering are positive in nature, across all four types of mind-wandering considered. However, their effect strength varies, with mind wandering relevant to chart appearance/topic boosting} affective engagement and positive emotional response ($\overline{\beta} = 0.51$), which can be attributed to its promoting deeper consideration of the visual elements or subject matter, fostering emotional engagement with the content. Mind wandering concerning chart data \anj{moderately boosts cognitive engagement ($\overline{\beta} = 0.39$), which may arise from its prompting deeper cognitive processing or interpretation.}

\noindent \textbf{Trust: }Complexity, Encodings, and Realism had direct negative effects over all the trust dimensions ($\overline{\beta} = -0.62\pm0.07$), while Annotation, Emphasis, and Text had direct positive effects ($\overline{\beta} = 0.58\pm0.13$). Structure and Comprehension showed weakly positive direct effects ($\overline{\beta} = 0.14\pm0.04$). The mind wandering frequency\footnote{\label{foot:ietmw}The indirect effects of the temporal distribution of mind wandering over all the paths, were in the expected direction (negative), \anj{but did not reach statistical significance; however, their inclusion did make the model perform better overall.}} was seen to partially mediate these paths ($\overline{\beta} = -0.27$), with a negative indirect effect. The indirect effect was strongest in the case of irrelevant mind wandering ($\overline{\beta} = -0.31$). We suggest that mind wandering causes users to view the information passively or superficially in turn leads to their failure to notice important details or connections within the visualization, reducing their overall trust in its validity and usefulness.

\noindent \textbf{Design Quality: }Annotation, Emphasis, Comprehension, and Structure had direct positive effects over design-quality dimensions ($\overline{\beta} = 0.58\pm0.11$), while Encodings, Realism, and Complexity had negative effects ($\overline{\beta} = -0.67$). Text was found to have a weaker negative effect ($\overline{\beta} = -0.36$), which can be attributed to its unique role in directing attention away from data encodings.  Mind wandering frequency was observed to partially mediate paths ($\overline{\beta} = -0.31$), with a negative indirect effect. This can be explained by the presence of design factors that increase cognitive load/disrupt attention (e.g., complexity, text), indirectly leading to higher mind wandering frequency, which in turn negatively impacted design quality. The indirect effect was weakest when mind wandering concerned a chart's appearance ($\overline{\beta} = -0.14$); we attribute this to users still interacting with the visualization's content even when momentarily distracted by its appearance.

\noindent \textbf{Adjectives: }Encodings and Realism had a direct positive effect on aesthetic appeal adjectives ($\overline{\beta} = 0.54$), while Text and Complexity had a direct negative effect ($\overline{\beta} = -0.58$). Annotation and Emphasis have a direct positive effect on communicative utility adjectives ($\overline{\beta} = 0.58$); Text and Comprehension had weaker positive effects ($\overline{\beta} = 0.41$), potentially due to their propensity to distract from the visual content or create visual noise. The remaining independent variables showed significant, but negligible effects for both types of adjectives ($\overline{\beta} = -0.04$). We found that mind wandering frequency$^3$ partially mediated paths for aesthetic appeal adjectives with a moderate, positive indirect effect ($\overline{\beta} = 0.37\pm0.08$), seen to increase when relevant to the chart's appearance ($\overline{\beta} = 0.45$). However, it also had a moderate negative impact on communicative utility adjectives ($\overline{\beta} = -0.39$), offset slightly when relevant to the chart's topic/data ($\overline{\beta} = -0.25$). Interestingly, the size of the indirect effects were greater than the direct effects, indicating that mind wandering accounted for a significant proportion of the relationship between design elements and elicited visualization descriptions. Thus, while mind wandering may enhance aesthetic appeal by fostering creativity or imaginative engagement with the visualization, it can also detract from its communicative utility by disrupting users' focus and attention.

\noindent \textbf{Short-Term Recall: }Text, Comprehension, Emphasis, and Annotation had a direct positive effect ($\overline{\beta} = 0.58$) on the accuracy and specificity of reported takeaways. Encodings, Realism, Text, and Structure had a direct positive effect on reports of the most salient visualization feature ($\overline{\beta} = 0.55$). The remaining independent variables showed significant, but negligible effects for recall ($\overline{\beta} = 0.05\pm0.01$). We also found that mind wandering frequency$^3$ partially mediated paths for takeaways ($\overline{\beta} = -0.34$) and saliency ($\overline{\beta} = -0.37$) reports, with a negative indirect effect. Saliency report accuracy was offset for apperance/data-relevant mind wandering ($\overline{\beta} = -0.31$). We posit that mind wandering may, disrupt the consolidation of memory traces related to the chart content, leading to decreased recall accuracy and specificity.

\noindent \textbf{Long-Term Recall: }Encodings, Structure, and Realism had a positive effect on long-term recall accuracy ($\overline{\beta} = 0.37$); the remaining independent variables showed negligible effects ($\overline{\beta} = 0.03\pm0.01$). Mind wandering frequency$^3$ partially mediated these paths in a weakly positive manner ($\overline{\beta} = 0.14$); however, the effect magnitude was extremely low. \anj{Irrelevant mind wandering is found to have a non-significant impact on long-term recall} ($\overline{\beta} = -0.05\pm0.01$), while relevant mind wandering (particularly concerning chart topic/appearance) has a moderate positive indirect effect ($\overline{\beta} = 0.35$, \anj{$p=0.048$}). Mind wandering episodes focused on relevant aspects of the visualization may facilitate semantic encoding, where users attribute personal meaning or significance to the information. Thus, a deeper level of processing can enhance memory consolidation by creating rich and interconnected memory traces that are more resistant to forgetting over time.


\textbf{\textit{Hence this hypothesis is partially supported for engagement, emotional response; fully supported for trust, design-quality, communicative utility adjectives, short-term recall; and not supported for aesthetic appeal adjectives and long-term recall.}}

\section{Conclusions and Future Work}


\anj{Our research highlights mind wandering as a dynamic measure of in situ viewer experience. Certain design elements, such as human recognizable objects, 3D, and photorealism, increase mind wandering, while non-text annotation, data/message redundancy, and high data ink ratio decrease it. This likely relates to their impact on cognitive load. Mind wandering also negatively mediates post-viewing measures. This suggests a need for deeper inquiry into the relationship between design elements, cognitive load, and memory consolidation.}


We note that our measure of mind wandering relies on self-report via key presses, consistent with practices in visual cognition literature~\cite{seli2013few,smith2018mind,thomson2014link}. However, self-reports may not always accurately distinguish mind wandering from attention lapses or task-related thoughts. \anj{Integrating eye tracking~\cite{pelagatti2020closer} in future studies could provide a more nuanced understanding.} As this research is exploratory, future studies can investigate the trade-offs between self-reporting and capturing mind wandering in a more natural user experience context.

The following conclusions summarize our high-level insights:

\textbf{\textit{Mind Wandering} functions as a dynamic measure of user experience.}
\textit{Mind Wandering} offers insights into cognitive and affective processes during interaction with visual content. It captures nuanced fluctuations in attention and engagement over time, providing deeper understanding of cognitive load, emotional responses, and information processing strategies, unlike static metrics that focus solely on observable behaviors or performance outcomes, post viewing. 

\textbf{Redundancy and Annotation reduce Mind Wandering.}
Visualizations with redundant information or clear annotations are less susceptible to mind wandering. Balancing data and message redundancy promotes clarity and organization, enhancing recognition and reducing mind wandering occurrences.  This balance can also facilitate comparisons across different parts of the data, resulting in a visually appealing and balanced design that promotes sustained engagement.

\textbf{\textit{Text} influences the type, frequency, and temporal distribution of mind wandering.}
Visualizations with extreme text volumes experience higher mind wandering frequencies. Moderate text volumes sustain attention without overwhelming users, aligning with practices in designing for ``slow analytics''. 

\anj{\textbf{\textit{Aesthetic novelty} increases the occurrence of topic-relevant mind-wandering.} 
The use of unique features like human recognizable objects and photorealism in charts, is linked to increased instances of topic-relevant mind wandering. This type of mind wandering often involves integrating external knowledge to form actionable insights, indicating that these features can stimulate deeper cognitive engagement with the visualization's subject matter. Similarly, elements that enhance aesthetic enjoyment, such as color, show weaker but similar effects. Conversely, elements like axes and gridlines decrease topic-relevant mind wandering and slightly increase irrelevant mind wandering.
}

\anj{\textbf{High perceived \textit{complexity} increases the occurrence of irrelevant mind-wandering.} 
Irrelevant mind-wandering is very common and triggered early on in the viewing period for charts that are not comprehensible ``at-a-glance". This behavior is heavily linked to high data volumes and visual density, especially if charts are sparsely annotated.} 

\anj{\textbf{\textit{Data-ink-ratio} offsets irrelevant mind-wandering}
The presence of high-data ink ratio in charts like isotypes, helps partially offset the negative effects of data volume and visual density, while still supporting high data dimensionality. The visual difficulties~\cite{hullman2011benefitting} introduced by the complex glyphs instead trigger data/topic-relevant mind wandering, though these continue to occur within the first 15 seconds of viewing.
}

\anj{
\textbf{\textit{Affective Engagement} can be enhanced by prompting relevant mind wandering.}
Embellishing familiar charts can enhance affective engagement by prompting relevant mind wandering. Familiarity offsets the impact of increased visual density, improving short-term recall and overall engagement. Further examination of how embellishments can be leveraged to promote redundancy can help inform more concrete guidelines on what is ``useful chart-junk''. 
}



\anj{Understanding how the brain processes visualizations dynamically is the first step toward effectively communicating crucial information intended by designers, ensuring sustained viewer focus and effective memory consolidation. To deepen insights into mind wandering's impact on visualization consumption, we aim to expand our database to include multi-panel visualizations.} Additionally, while we do observe broad general trends between engagement and mind wandering, \anj{it's possible that self-reports of mind-wandering exacerbates participants' lack of engagement, as the self-reports constitute a secondary task being performed. This stems from the dependence of self-reports on participants' continued awareness of their internal state of attention, without any prompting. Similarly, the use of the key-press paradigm requires that participants accurately categorize mind wandering instances for self-reports, leading to a greater allocation of working memory to this process. Investigating whether our paradigm impacts the frequency of self-reports seen over the course of the study is out of scope, due to the diversity of stimuli used. However, future work aims to explore alternative methods for eliciting mind wandering, including physiological monitoring, to better reflect real-world visualization consumption paradigms. Additionally, investigating variations in observation time periods could reveal how mind wandering dynamics affect change detection, and the memorability/retention of visual elements. Future studies also intend to assess the generalizability of findings across diverse populations, including older adults and non-native English speakers, who may have different executive control capabilities. Lastly, we plan on designing interaction techniques to foster relevant mind wandering and increase emotional engagement while maintaining analytical focus.}


\anj{By understanding how viewers dynamically consume visualizations, future studies can address critical questions about systematically refining design and presentation to enhance user engagement and comprehension. Investigating the impact of low-level visual elements on mind wandering, as well as mind wandering's influence on post-viewing measures, can inform control strategies in future experiments and guide the creation of guidelines for crafting impactful, holistic visualizations. }

\section{Appendices}
\label{sec:appendices_inst}

\section*{Supplemental Materials}
\label{sec:supplemental_materials}

\anj{Supplemental materials are available  at \url{https://osf.io/h5awt/}, released under a CC BY 4.0 license.}
These include:
(1) the aggregate data for collected measures, (2) stimuli used in the study with metadata, (3) SEM Analysis results (factor loadings, chi-squared, regression coefficients, significance, Cohen's f$^2$), (4) demographic data from Experiments, and (5) a full version of this paper with all appendices.

\section*{Figure Credits}
\label{sec:figure_credits}
\anj{Figure 1 -- Is it a bird ? by Annabelle Rincon, Tableau Public, 2021.} Figure 6 -- Nigel Holmes, Time Magazine, 1982.

\bibliographystyle{abbrv-doi-hyperref}

\bibliography{00_template}

\appendix 

\end{document}